\documentstyle[aps,prl,twocolumn,graphics]{revtex}

\begin{document}
\draft
\wideabs{
\title{Classical Analog of Electromagnetically Induced 
Transparency}
\author{C. L. Garrido Alzar$^{1}$, M. A. G. Martinez$^{2}$, and 
P. Nussenzveig$^{1}$} 

\vskip 0.3cm 

\address{$^{1}$ Instituto de F\'\i sica, Universidade de S\~ao Paulo, 
Caixa Postal 66318, 05315-970 S\~ao Paulo, SP, Brazil \\ [0.2cm] 
$^{2}$ P\'os-gradua\c c\~ao em Engenharia El\'etrica, Universidade 
Presbiteriana Mackenize, Rua da Consola\c c\~ao 896, \\ 01302-000, S\~ao 
Paulo, SP, Brazil}
\date{April 2001}
\maketitle

\begin{abstract}
We present a classical analog for Electromagnetically 
Induced Transparency (EIT). In a system of just two coupled 
harmonic oscillators subject to a harmonic driving force 
we can reproduce the phenomenology observed in EIT. We describe 
a simple experiment performed with two linearly coupled $RLC$ 
circuits which can be taught in an undergraduate laboratory class. 
\end{abstract}
\pacs{ }
}

\section{Introduction}
\label{sec:intro} 
\indent 

Imagine a medium which strongly absorbs a light beam of a 
certain frequency. Add a second light source, with a 
frequency that would also be absorbed by the medium, and 
the medium becomes transparent to the first light beam. 
This curious phenomenon is called Electromagnetically 
Induced Transparency, or simply EIT~\cite{harris}. It usually takes 
place in vapors of three-level atoms. The light sources are 
lasers which drive two different transitions with one 
common level. Most authors attribute the effect to quantum 
interference in the atomic medium, involving two 
indistinguishable paths leading to the common state. 
The dispersive properties of the medium are also significantly 
modified as has been recently demonstrated with the impressive 
reduction of the group velocity of a light pulse to only 
17~m/s~\cite{hau,kash,budker} and the ``freezing'' of a 
light pulse in an atomic medium~\cite{hau2,lukin}. 

In this paper we develop a totally classical analog for 
Electromagnetically Induced Transparency and present a 
simple experiment which can be carried out in an 
undergraduate physics laboratory. The stimulated resonance 
Raman effect has already been modeled classically in a 
system of three coupled pendula by 
Hemmer and Prentiss~\cite{hemmprent88}. Even though many 
aspects of EIT are already present in the stimulated 
Raman effect, as can be seen in~\cite{hemmprent88}, at that 
time EIT had not yet been observed~\cite{harris} and the 
dispersive features were not considered. 
Our model involves only two oscillators with linear 
coupling. The experiment is performed with $RLC$ circuits. 
The interest of such an experiment and purpose of this paper 
is to enable undergraduate students to develop physical 
intuition on coherent phenomena which occur in atomic systems. 

\newpage
\section{Theoretical Model}
\label{sec:theory} 
\indent 

We will focus our attention on the simulation of EIT in material 
media composed of three-level atoms in the so-called $\Lambda$ 
configuration interacting with two laser fields as shown in Fig.~1. 
The quantum states $|1\rangle$ and $|2\rangle$ represent the two ground 
states of the atom, and the state $|0\rangle$ is the excited atomic level. 

\begin{figure}
\centering \resizebox{7.5cm}{!}{\includegraphics*{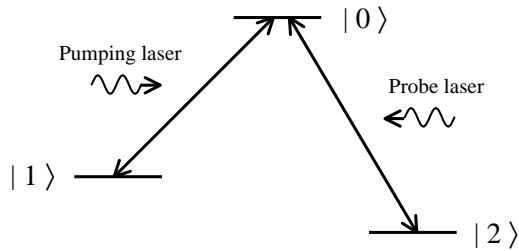}}
\caption{Energy diagram of a three-level $\Lambda$-type atom interacting 
with two laser beams, coupling the two ground states to a common excited 
atomic state.}
\label{fig1}
\end{figure}

The laser field coupled to the atomic transition between the states $|1\rangle$ 
and $|0\rangle$ will be called ``pumping (or pump) laser'', and the laser 
coupled to the transition between the states $|2\rangle$ and $|0\rangle$ will 
be the ``probe laser''. A typical experiment consists of scanning the frequency 
of the probe laser and measuring its transmitted intensity. In the absence of 
the pump laser, one observes a standard absorption resonance profile. Under 
certain conditions, the addition of the pump laser prevents absorption in a 
narrow portion of the resonance profile, and the transmitted intensity as a 
function of the probe frequency presents a narrow peak of induced transparency. 

The effect depends strongly on the pump beam intensity. Typically the 
pump laser has to be intense so that the Rabi frequency $\Omega_1$ associated to 
the transition from state $|1\rangle$ to $|0\rangle$ is larger than all 
damping rates present (associated to spontaneous emission from the 
excited state and other relaxation processes). One of the effects 
of this pump beam is to induce an AC-Stark splitting of the excited 
atomic state. The probe beam will therefore couple state $|2\rangle$ to 
two states instead of one. If the splitting (which varies linearly with 
the Rabi frequency $\Omega_1$) is smaller than the excited state 
width, the two levels are indistinguishable and one expects quantum 
interference in the probe absorption spectrum. As the Rabi frequency 
$\Omega_1$ increases, the splitting becomes more pronounced and 
indistinguishability is lost. The absorption spectrum becomes a doublet 
called the Autler-Townes doublet. We will keep this image in mind to 
present a classical system with the same features. 

\subsection{EIT-like phenomena with masses and springs} 
\label{subsec:eitmec} 
\indent 

We will model our atom as a simple harmonic oscillator, consisting 
of a particle 1, of mass $m_1$, attached to two springs with spring 
constants $k_1$ and $K$ (see Fig.~2). The spring with constant 
$k_1$ is attached to a wall, while the other spring is attached to a second 
particle of mass $m_2$ and initially kept immobile at a fixed position. 
Particle 1 is also subject to a harmonic force 
${\cal F}_s=F e^{-i\ (\omega_{s} t + \phi_s)}$. 
If we analyze the power transferred from the harmonic source to particle 
1, as a function of frequency $\omega_s$, in this situation we will 
observe the standard resonance absorption profile discussed above 
(peaked at frequency $\omega_{1}^{2}=(k_{1}+K)/m$). If we now allow 
particle 2 to move, subject only to the forces from the spring of 
constant $K$ and a third spring of constant $k_2$, attached to a wall 
(see Fig.~2), this profile is modified. As we shall see, 
depending on the spring constant $K$, we can observe a profile 
presenting features similar to EIT evolving to an 
Autler-Townes-like doublet (as a matter of fact, this doublet is simply 
a normal-mode splitting). For simplicity, we will consider the 
situation in which $k_1=k_2=k$ and $m_1=m_2=m$. 

The physical analogy between our model and the three-level atom 
coupled to two light fields is simple. As mentioned, the atom is 
modeled as oscillator 1 (particle 1), with its resonance frequency 
$\omega_1$. Since we chose $k_1=k_2=k$ and $m_1=m_2=m$, we have 
the analog of a degenerate $\Lambda$ system. The pump field is 
simulated by the coupling of oscillator 1 to a second oscillator 
via the spring of constant $K$ (reminding us of the quantized 
description of the field, in terms of harmonic oscillators). 
The probe field is then modeled by the harmonic force acting on 
particle 1. 

\begin{figure}
\centering \resizebox{7.5cm}{!}{\includegraphics*{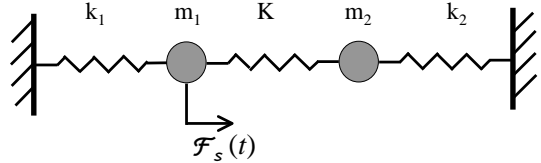}}
\caption{Mechanical model used to simulate EIT.}
\label{fig2}
\end{figure}

In order to provide a quantitative description of the system 
we write the equations of motion of particles 1 and 2 in terms 
of the displacements $x_1$ and $x_2$ from their respective 
equilibrium positions: 

\begin{eqnarray}
 \ddot{x_{1}}(t)+\gamma_{1}\dot{x_{1}}(t)+\omega^{2}x_{1}(t)-
 \Omega_{r}^{2}x_{2}(t) = \frac{F}{m} e^{-i\ \omega_{s} t} \nonumber \\ [0.2cm]
 \ddot{x_{2}}(t)+\gamma_{2}\dot{x_{2}}(t)+\omega^{2}x_{2}(t)-
 \Omega_{r}^{2}x_{1}(t) = 0 \;,
\label{eq15}
\end{eqnarray} 

\noindent
where we set $\phi_{s}=0$ for the probe force without loss of 
generality. We also defined $\Omega_{r}^{2}=K/m$, the frequency 
associated with the coherent coupling between 
the pumping oscillator and the oscillator modeling the atom, 
$\gamma_{1}$ the friction constant associated to the energy 
dissipation rate acting on particle 1 (which simulates 
the spontaneous emission from the atomic excited state), and 
$\gamma_{2}$, the energy dissipation rate of the pumping transition. 

Since, we are interested in the power absorbed by particle 1 from 
the probe force, we will seek a solution for $x_{1}(t)$. Let us suppose 
that $x_{1}(t)$ is of the form 

\begin{equation}
 x_{1}(t) = N e^{-i\ \omega_{s} t} \;,
\label{eq16}
\end{equation} 

\noindent
where $N$ is a constant. After taking a similar expression 
for $x_{2}(t)$ and substituting in eq.~(\ref{eq15}), we find 

\begin{equation}
 x_{1}(t)=\frac{(\omega^{2}-\omega_{s}^{2}-
 i \gamma_{2} \omega_{s}) F e^{-i\ \omega_{s} t}}
 {m[(\omega^{2}-\omega_{s}^{2}-i \gamma_{1} \omega_{s})
 (\omega^{2}-\omega_{s}^{2}-i \gamma_{2} \omega_{s})-\Omega_{r}^{4}]} \;.
\label{eq17}
\end{equation}

Now, computing the mechanical power $P(t)$ absorbed by particle 1 from the probe 
force ${\cal F}_s$, 

\begin{equation}
 P(t)=F e^{-i\ \omega_{s} t} \dot{x_{1}}(t) \;, 
\label{eq18}
\end{equation}

\noindent
we find for the power absorbed during one period of oscillation of the 
probe force 

\begin{equation}
 P_{s}(\omega_{s})=-\frac{2 \pi i F^{2} \omega_{s}
 (\omega^{2}-\omega_{s}^{2}-i \gamma_{2} \omega_{s})}
 {m[(\omega^{2}-\omega_{s}^{2}-i \gamma_{1} \omega_{s})
 (\omega^{2}-\omega_{s}^{2}-i \gamma_{2} \omega_{s})-\Omega_{r}^{4}]} \;.
\label{eq19}
\end{equation}

In Fig.~3 we show the real part of $P_{s}(\omega_{s})$, for six 
different values of the coupling frequency $\Omega_{r}$ expressed in 
frequency units. These curves were obtained using the values 
$\gamma_{1}=0.1\times 10^{-6}$, $\gamma_{2}=0.4\times 10^{-1}$, and 
$\omega_0 =\sqrt{k/m}=2.0$, all expressed in the 
same frequency units. The amplitude $F/m$ was taken equal to $0.1$ 
force per mass units.

For $\Omega_{r}=0$ we have a typical absorption profile, with a maximum 
probe power absorption for $\delta=0$, being $\delta$ the detuning between 
the probe and the oscillator frequencies ($\delta=\omega_{s}-\omega$). 
Incrementing the value of $\Omega_{r}$ to 0.1, we observe the appearance 
of a narrow dip in the absorption profile of the probe power. This zero 
absorption at the center frequency of the profile is an evidence of a 
destructive interference between the normal modes of oscillation of the 
system~\cite{foot}. A further increment 
of the coupling frequency leads to the apparition of two peaks in the 
probe power absorption profile, that are clearly separated for 
$\Omega_{r}=0.5$. This effect in atomic systems is called the AC-Stark 
splitting or Autler-Townes doublet. 

\begin{figure}
\centering \resizebox{7.5cm}{!}{\includegraphics*{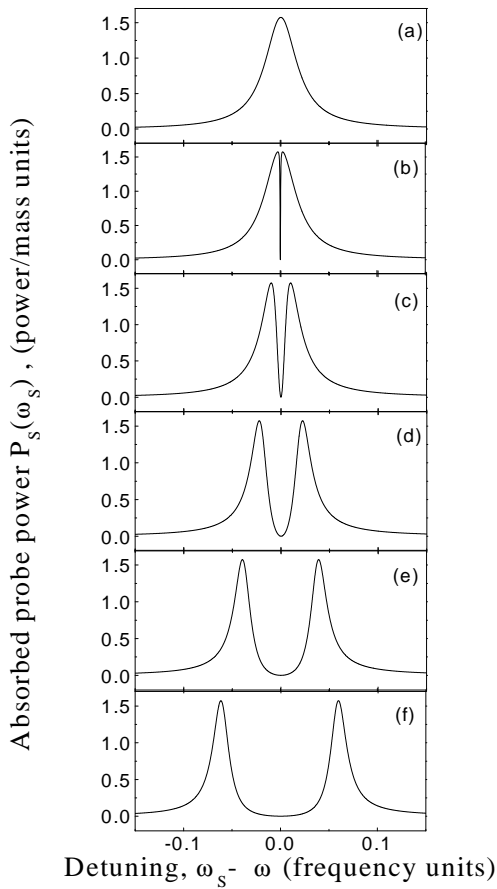}}
\caption{Frequency dependence of the absorption of the probe energy by 
particle 1. The values of $\Omega_r$ in each case are (a) 0.0 , (b) 0.1 , 
(c) 0.2 , (d) 0.3 , (e) 0.4 , and (f) 0.5 , all expressed in frequency 
units.}
\label{fig3}
\end{figure}

An important fact to be pointed out is that the dissipation rate 
$\gamma_{2}$, associated to the energy loss of the pumping oscillator, must 
be much smaller than $\gamma_{1}$ in order to achieve a regime of 
coherent driving of particle 1's oscillation. In other words, there should 
be no significant increase of dissipation in the system by adding the 
pumping oscillator. In the case of EIT~\cite{li}, the condition analogous 
to $\gamma_2 \ll \gamma_1$ is that the transfer rate of population 
from the atomic ground state $|1\rangle$ to $|2\rangle$ should be negligible 
(see Fig.~1). In both situations, the violation of this condition obstructs 
the observation of the induced transparency. 
 
Another important result reproduced with this mechanical model is the 
dispersive behavior of the mass oscillator used to simulate the atom. 
The dispersive response is contained in the real part of the frequency 
dependence of the amplitude of oscillation $x_{1}(t)$ of particle 1, given 
by eq.~(\ref{eq17}). In Fig.~4 we plot this quantity for 
$\Omega_{r}=0.1$. This value of $\Omega_{r}$ corresponds to the situation 
when the induced transparency is more pronounced and, as we can see from 
Fig.~4, the dispersion observed, in the frequency interval where 
we have the induced transparency, is normal and with a very steep slope. 

\begin{figure}
\centering \resizebox{7.5cm}{!}{\includegraphics*{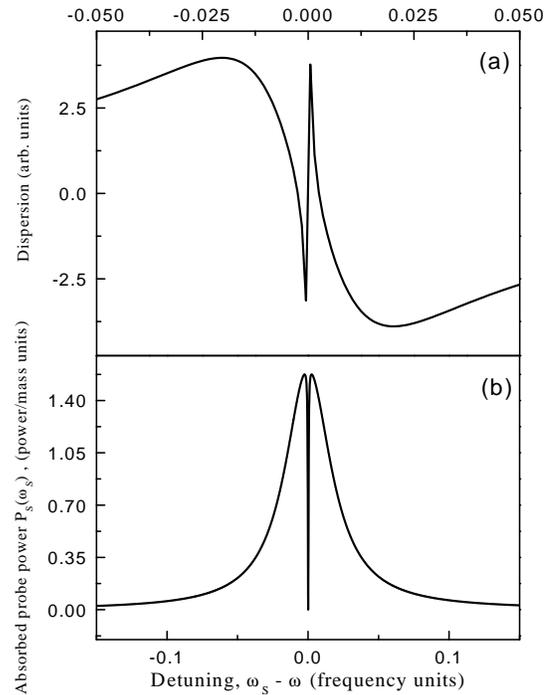}}
\caption{Dispersive (a) and absorptive (b) responses of the probe power 
transferred to particle 1, for $\Omega_r=0.1$ frequency units.}
\label{fig4}
\end{figure}

\noindent
This result coincides with that reported before in an experimental realization of 
electromagnetically induced transparency~\cite{xiao}. It is also important 
to point out that this very steep normal dispersion is responsible for the 
recently observed slow propagation of light (slow light)~\cite{hau,kash,budker}, 
and the ``freezing'' of a light pulse in an atomic medium~\cite{hau2,lukin}. 
It should therefore be possible to observe such propagation effects considering 
absorption in a medium consisting of a series of mechanical `atom-analogs'. 

This theoretical model is not the only classical 
one to simulate EIT-like phenomena. As mentioned, in this model 
the pump field is replaced by a harmonic oscillator, simulating 
a quantum-mechanical description. In most theoretical descriptions 
of EIT in atomic media, pump and probe fields are classical. 
The mechanical analog to this description would 
then involve only one oscillator (one particle of mass $m$). 
In this analysis, it becomes apparent that EIT arises directly from a 
destructive interference between the oscillatory forces 
driving the particle's movement. In order to keep the text simple, 
we have chosen to not present this description here. 
Furthermore, it is not related to the experimental results 
presented below. 

\section{EIT-like phenomena in coupled $RLC$ circuits: a simple 
undergraduate laboratory experiment} 
\label{sec:eitrlc_exp} 
\indent 

An experiment to observe the predictions of the previous section, 
although possible, would not be straightforward. Instead, we 
decided to use the well-known analogy between mechanical oscillators 
and electric circuits to perform a simple experiment. The electric 
analog to the system of Fig.~2, is the circuit shown in 
Fig.~5, where the circuit mesh composed by the inductor 
$L_{1}$ and the capacitors $C_{1}$ and $C$ simulates the pumping oscillator, and 
the resistor $R_{1}$ determines the losses associated with that oscillator. The 
atom is modeled using the resonant circuit formed by the 
inductor $L_{2}$, the capacitors $C_{2}$ and $C$, and the resistor $R_{2}$ 
represents the spontaneous radiative decay from the excited level. 
The capacitor $C$, belonging to both circuit meshes, models the coupling 
between the atom and the pumping field, and determines the Rabi frequency 
associated to the pumping transition. In this case, the probe field is simulated 
by the frequency-tunable voltage source $V_{S}$.

\begin{figure}
\centering \resizebox{7.5cm}{!}{\includegraphics*{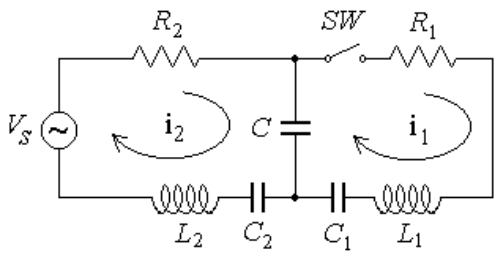}}
\caption{Electric circuit employed to investigate the induced 
transparency.}
\label{fig5}
\end{figure}

The circuit mesh used to model the atom has only one resonance frequency 
representing the energy of the atomic excited level. That is to say, the 
probability of excitation of this circuit will be maximal when 
the applied harmonic force is on resonance. Since in this case 
we have two possible paths to accomplish this excitation, we are dealing with 
the analog of a three-level atom in the $\Lambda$ configuration. Namely, 
the oscillator corresponding to the `atom-analog' can be excited directly 
by the applied voltage $V_s$ or by the coupling to the pumping oscillator. 

Here again the induced transparency is investigated analyzing the frequency 
dependence of the power $P_{2}(\omega_{s})$ transferred from the voltage source 
$V_{S}$ to the resonant circuit $R_{2}L_{2}C_{e2}$, 

\begin{equation}
 P_{2}(\omega_{s})=\Re\{{\cal V}_{S}{\cal I}_{2}^{*}\} \;,
\label{eq20}
\end{equation}

\noindent
where ${\cal V}_{S}$ and ${\cal I}_{2}$ are respectively the complex 
representations of $V_{S}$ and $i_{2}(t)$, and the equivalent capacitor 
$C_{e2}$ is the series combination of $C$ and $C_{2}$: 

\begin{equation}
 C_{e2}=\frac{C C_{2}}{C+C_{2}} \;.
\label{eq21}
\end{equation}

Setting $L_{1}=L_{2}=L$ ($m_1=m_2=m$ in the mechanical model) and writing the 
equations for the currents $i_{1}(t)=\dot{q_{1}}(t)$ and 
$i_{2}(t)=\dot{q_{2}}(t)$ shown in Fig.~5, we find the following 
system of coupled differential equations for the charges 
$q_{1}(t)$ and $q_{2}(t)$: 

\begin{eqnarray}
 \ddot{q_{1}}(t)+\gamma_{1}\dot{q_{1}}(t)+\omega_{1}^{2}q_{1}(t)-
 \Omega_{r}^{2}q_{2}(t) = 0 \nonumber \\
 \ddot{q_{2}}(t)+\gamma_{2}\dot{q_{2}}(t)+\omega_{2}^{2}q_{2}(t)-
 \Omega_{r}^{2}q_{1}(t) = \frac{V_{S}(t)}{L_{2}} \;.
\label{eq22}
\end{eqnarray} 

\noindent
Here $\gamma_i = R_i/L_i$ ($i=1,2$), $\omega_i^2 = 1/(L_i C_{ei})$, 
and $\Omega_r^2 = 1/(L_2 C)$. These equations coincide with the 
eqs.~(\ref{eq15}) using the correspondences shown in Table~1 and 
$\omega_1=\omega_2$. Therefore, both models describe the same physics.

\begin{table}[htb]
\begin{center}
\begin{tabular}{|c|c|}
\hline
Mechanical model & Electrical model \\
\hline \hline
$\gamma_{1}$ ($\gamma_{2}$) & $R_{1}/L_1$ ($R_{2}/L_2$) \\
$m_1$ ($m_{2}$) & $L_{1}$ ($L_{2}$) \\
$k_{1}$ ($k_{2}$) & $1/C_{1}$ ($1/C_{2}$) \\
$K$ & $1/C$ \\
$x_{1}$ ($x_{2}$) & $q_{1}$ ($q_{2}$) \\
$(F/m) \, e^{-i\ \omega_{s}t}$ & ${\cal V}_{S}(t)/L_{2}$ \\
\hline
\end{tabular}
\end{center}
\label{tab1}
\caption{Correspondences between the mechanical and electrical parameters.}
\end{table}

Once the current $i_{2}(t)$ (or ${\cal I}_{2}$) is known, the expression that 
determines the power $P_{2}(\omega_{s})$ as a function of the frequency 
$\omega_{s}$ of the applied voltage, when the switch $SW$ in Fig. 5 is 
closed, is given by 

\begin{equation}
 P_{2}(\omega_{s})=\frac{p1(\omega_{s})}
 {p1(\omega_{s})^{2}+p2(\omega_{s})^{2}}|A_{S}|^{2} \;, 
\label{eq23}
\end{equation}

\noindent 
where 

\begin{equation}
 p1(\omega_{s})=R_{2}+\frac{R_{1}/(\omega_{s} C)^{2}}
 {R_{1}^{2}+(\omega_{s}L_{1}-1/(\omega_{s}C_{e1}))^{2}}
\label{eq231}
\end{equation}

\begin{eqnarray}
 p2(\omega_{s}) &=& (\omega_{s} L_{2}-1/(\omega_{s}C_{e2}))- 
\nonumber \\ [0.15cm]
 & & \frac{(1/(\omega_{s} C)^{2})(\omega_{s}L_{1}-1/(\omega_{s}C_{e1}))}
 {R_{1}^{2}+(\omega_{s}L_{1}-1/(\omega_{s}C_{e1}))^{2}} \;,
\label{eq232}
\end{eqnarray} 
being $C_{e1}$ the equivalent capacitor given by the series combination of 
$C$ and $C_{1}$ and $A_{S}$ represents the amplitude of the applied voltage. 
On the other hand, when the switch is open we have 
\begin{equation}
 P_{2}(\omega_{s})=\frac{R_{2}}{R_{2}^{2}+
 (\omega_{s} L_{2}-1/(\omega_{s}C_{e2}))^{2}}|A_{S}|^{2} \;.
\label{eq24}
\end{equation}

There are many ways of measuring the power $P_{2}(\omega_{s})$. We have chosen 
to measure the current flowing through the inductor $L_{2}$, which 
has the same frequency dependence as $P_{2}(\omega_{s})$. We actually measure 
the voltage drop across the inductor $L_{2}$ and integrate it to find a 
voltage proportional to the current flowing through the inductor. This 
voltage is an oscillatory signal at the frequency $\omega_s$. We are 
interested in the amplitude of this signal, which we read off an 
oscilloscope. 

In Fig.~6 are presented the experimental amplitudes measured, corresponding to 
four different values of the coupling capacitor $C$. For each value of the 
coupling capacitor a measurement was made in two situations: with the 
switch open (open square) and with the switch closed (open circle). In 
Table~2 we present the specifications of the electronic components 
used in the experiment. 

\begin{table}[tbh]
\begin{center} 
\begin{tabular}{|c|c|}
\hline
Electronic component & Value \\
\hline \hline
$R_{1}$ & $0$ Ohms \\
$R_{2}$ & $51.7$ Ohms \\
$L_{1}$ & $1000$ $\mu$~H \\
$L_{2}$ & $1000$ $\mu$~H \\
$C_{1}$ & $0.1$ $\mu$~F \\
$C_{2}$ & $0.1$ $\mu$~F \\
\hline
\end{tabular}
\end{center}
\caption{Specifications of the electronic components used in the experiment.}
\label{tab2}
\end{table}

\begin{figure}
\centering \resizebox{7.5cm}{!}{\includegraphics*{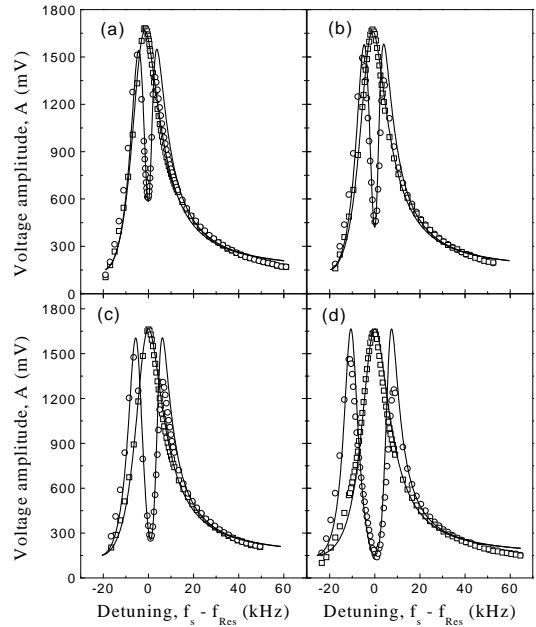}}
\caption{Power transferred to the circuit 
$R_2 L_2 C_{e2}$ as a function of the frequency $\omega_s$ for different 
values of the coupling capacitor $C$. The values of $C$ in each curve are 
(a) $C=0.196\mu$~F, with resonance frequency f$_{\mbox{\scriptsize Res}} = 
\omega_2 / 2\pi = 20.0$~kHz, (b) $C=0.150\mu$~F, f$_{\mbox{\scriptsize Res}} 
= 19.5$~kHz, (c) $C=0.096\mu$~F, f$_{\mbox{\scriptsize Res}} = 21.5$~kHz, 
(d) $C=0.050\mu$~F, f$_{\mbox{\scriptsize Res}} = 26.5$~kHz. The open squares 
correspond to measurements made with the switch $SW$ open (see Fig.~5), 
which is equivalent to turning the pump off. The open circles correspond to 
measurements made with the switch closed. The solid curves are a comparison 
to the theory presented in \S~\ref{sec:eitrlc_exp}. The evolution from EIT 
to the Autler-Townes regime is clearly observed.}
\label{fig6}
\end{figure}

As we can see from Fig.~6, in the open switch configuration (no 
pumping) we have a maximum coupling of electrical power from the voltage source 
$V_{S}$ to the resonant circuit $R_{2}L_{2}C_{e2}$ at the resonant frequency 
(zero detuning). When the switch is closed, that is to say, the pumping circuit 
(pumping force) is acting on the resonant circuit $R_{2}L_{2}C_{e2}$, we have a 
depression of the absorption of the electrical power of the voltage source at 
zero detuning. This fact corresponds to the transparent condition, and is 
more pronounced when the value of the coupling capacitor is reduced, 
corresponding to an increment of the Rabi frequency of the pumping field in 
electromagnetically induced transparency studied in atomic systems.

Here, as in the experiments with atoms and light, the observed transparency 
can be interpreted as a destructive interference. In this case, 
the inteference is between the power given by the 
voltage source to the resonant circuit $R_{2}L_{2}C_{e2}$, and the power 
transferred to the same circuit from the other oscillator representing the 
pumpimg force. As explained in \S\ref{subsec:eitmec}, it can also 
be viewed as an interference between the two possible excitation paths 
corresponding to the normal modes of the coupled oscillators. 
For the minimum value of $C$ used in our experiment (see 
Fig.~6(d)), we observe two absorption peaks, as the classical 
analog of the Autler-Townes effect or AC-Stark splitting, which 
correspond to the splitting of these normal modes. We 
should also point out that the smallest coupling value we used 
(Fig.~6(a)) did not lead to an infinitely narrow transparency 
peak as would be expected from the value $R_1=0$. This is probably 
due to internal ``residual'' series resistances of the components we used. 

The solid lines shown in Fig.~6 represent the theoretical 
results obtained using eqs.~(\ref{eq23}) or (\ref{eq24}). These curves are 
not intended to fit exactly the experimental data since our measurements are 
affected by the frequency response of the integrator used to derive the voltage 
proportional to the current in the inductor $L_{2}$. It is not our purpose 
to analyze in detail the deviations of our system from the ideal model 
proposed. 

It is also possible to measure the dispersive response of our `atom-analog'. 
One has to analyze the relative phase between the oscillating current 
flowing through inductor $L_2$ and the phase of the applied voltage. This 
measurement is best if performed with a lock-in amplifier. Since this 
equipment is not usually available in undergraduate laboratories, we 
prefer to describe a simple procedure which does not lead to a real 
measurement to be compared with the theoretical prediction (Fig.~4). 
We observe the oscillatory signal on the oscilloscope corresponding 
to the current flowing through the inductor, with the trigger signal 
from the voltage source $V_S(t)$. The phase of the sinusoidal signal is 
observed to ``jump'' as we scan $\omega_s$ across the absorption resonance 
and transparency region. If we scan $\omega_s$ increasing its value, we 
observe three abrupt phase variations, with the intermediate one being 
opposite to the other two. This is exactly what we would expect from 
Fig.~4. 

\section{Conclusion}
\label{sec:conclu} 
\indent 

We have shown that EIT can be modeled in a totally 
classical system. Our results extend the results of 
Hemmer and Prentiss~\cite{hemmprent88} to EIT-like 
phenomena, with the use of only two coupled oscillators instead 
of three. We also deal with the dispersive response of 
the classical oscillator used to model the three-level 
atom. We performed an experiment with coupled $RLC$ 
circuits and observed EIT-like signals and the classical analog 
of the Autler-Townes doublet as a function of the coupling between 
the two $RLC$ circuits. The experiment can be 
performed in undergraduate physics laboratories and should 
help form physical intuition on coherent phenomena which 
take place in atomic vapors. 

\section*{Acknowledgments} 
\indent 

We acknowledge financial support from the Brazilian agencies FAPESP, 
CAPES, and CNPq. 



\begin{thebibliography}{99}

\bibitem{harris} See, for example, S. E. Harris, ``Electromagnetically 
Induced Transparency'', Phys. Today \textbf{50} (7), 36-42 (1997).

\bibitem{hau} L. V. Hau, S. E. Harris, Z. Dutton, and C. H. Behroozi, 
``Light speed reduction to 17 metres per second in an ultracold atomic gas'', 
Nature \textbf{397}, 594-598 (1999). 

\bibitem{kash} M. M. Kash, V. A. Sautenko, A. S. 
Zibrov, L. Hollberg, G. R. Welch M. D. Lukin, Y. Rostovtsev, E. S. Fry, and 
M. O. Scully, ``Ultraslow Group Velocity and Enhanced Nonlinear Optical 
Effects in a Coherently Driven Hot Atomic Gas'', Phys. Rev. Lett. 
\textbf{82}, 5229-5232 (1999). 

\bibitem{budker} D. Budker, D. F. Kimball, S. M. Rochester, 
and V. V. Yashchuk, ``Nonlinear Magneto-optics and Reduced Group Velocity of 
Light in Atomic Vapor with Slow Ground State Relaxation'', Phys. Rev. Lett. 
\textbf{83}, 1767-1770 (1999).

\bibitem{hau2} C. Liu, Z. Dutton, C. H. Behroozi, and L. V. Hau, 
``Observation of coherent optical information storage in an atomic 
medium using halted light pulses'', Nature \textbf{409}, 490-493 (2001). 

\bibitem{lukin} D. F. Phillips, A. Fleischhauer, A. Mair, R. L. Walsworth, 
and M. D. Lukin, ``Storage of Light in Atomic Vapor'', Phys. Rev. Lett. 
{\bf 86}, 783-786 (2001). 

\bibitem{hemmprent88} 
P. R. Hemmer and M. G. Prentiss, ``Coupled-pendulum model of the 
stimulated resonance Raman effect'', J. Opt. Soc. Am. B {\bf 5}, 
1613-1623 (1988). 

\bibitem{foot} As explained, the induced transparency results from 
the existence of two possible paths for the absorption of the probe energy 
to excite the oscillation of particle 1. We can see these paths in 
eq.~(\ref{eq19}) rewriting it as a superposition of the normal modes 
of oscillation of the coupled oscillators.

\bibitem{li} Y.-q. Li, and M. Xiao. ``Observation of quantum interference 
between dressed states in an electromagnetically induced transparency'', 
Phys. Rev. A \textbf{51}, 4959-4962 (1995).

\bibitem{xiao} M. Xiao, Y.-q. Li, S.-z. Jin, and J. Gea-Banacloche, 
``Measurement of Dispersive Properties of Electromagnetically Induced 
Transparency in Rubidium Atoms'', Phys. Rev. Lett. \textbf{74}, 666-669 
(1995). 

\end{thebibliography}
\end{document}